\documentstyle[sprocl,epsfig]{article}

\def\Journal#1#2#3#4{{#1} {\bf #2}, #3 (#4)}

\def\PRD{{\em Phys. Rev.} D}
\def\ZPC{{\em Z. Phys.} C}
\def\MPL{{\em Mod. Phys. Lett.}  A}

\def\be{\begin{equation}}
\def\ee{\end{equation}}
\def\bea{\begin{eqnarray}}
\def\eea{\end{eqnarray}}

\begin{document}

\title{SPIN  EFFECTS  IN DIFFRACTIVE $J/\Psi$  AND $Q \bar Q$
       LEPTOPRODUCTION}

\author{ S.V.Goloskokov}

\address{Bogoliubov Laboratory of
Theoretical
  Physics, Joint Institute for\\  Nuclear Research, Dubna
  141980, Moscow
  region, Russia.\\E-mail: goloskkv@thsun1.jinr.dubna.su}


\maketitle\abstracts{We  study the dependencies of spin 
asymmetries in diffractive leptoproduction on the spin structure 
of the pomeron coupling.  It is shown  that the $A_{ll}$ 
asymmetry in diffractive processes is proportional to the 
fraction  of the initial proton momentum $x_p$ carried off by the 
pomeron. The resulting asymmetries decrease with increasing energy for  
diffractive  $J/\Psi$ production and are energy independent for 
diffractive $Q \bar Q$ leptoproduction. The connection of these 
asymmetries with the non-forward gluon distribution in the proton 
is discussed. }

The study of nature of the pomeron becomes again popular now due 
to the progress in investigation of diffractive reactions at 
HERA. The diffractive  $J/\Psi$ and  $Q \bar Q$ leptoproduction 
is a keystone of this problem.  The pomeron  is a vacuum 
$t$-channel exchange that contributes to high-energy diffractive 
processes.  The hadron-hadron scattering amplitude determined by 
the pomeron exchange can be written in the form
$T(s,t)^{A,B}={\rm i} I\hspace{-1.6mm}P(s,t) 
V_{AI\hspace{-1.1mm}P} \otimes V^{BI\hspace{-1.1mm}P} $,   
where $I\hspace{-1.6mm}P$ is a "bare" pomeron contribution, and
$V_{AI\hspace{-1.1mm}P}$ and $V^{BI\hspace{-1.1mm}P}$ are the
pomeron couplings with particles $A$ and $B$, respectively.

The spin structure of the pomeron coupling is an open problem up 
to now. When the gluons from the pomeron couple to a single quark 
in the hadron, a simple matrix structure of the pomeron vertex $V_{h 
I\hspace{-1.1mm}P}^{\mu} =\beta_{h I\hspace{-1.1mm}P} 
\gamma^{\mu}$ appears.
This standard coupling leads to spin-flip effects decreasing  
with increasing energy like $1/s$. The complicated spin structure of the 
pomeron coupling can be connected with the nonperturbative 
structure of the proton. Really, in a QCD--based model in which 
the proton is viewed as being composed of a quark and a diquark,  
the following structure of the proton coupling with a two-gluon 
system has been found
\begin{equation}
V_{pI\hspace{-1.1mm}P}^{\mu\nu}(p,r)= 4 p^{\mu} p^{\nu} A(r) 
+(\gamma^{\mu} p^{\nu} +\gamma^{\nu} p^{\mu}) B(r). 
\label{ver}
\end{equation}
Here $r$ is the momentum transfer. The term proportional to $B(r)$ 
represents the standard pomeron coupling that leads to the 
non-flip amplitude. The $A(r)$ contribution is determined by the 
 effects of vector diquarks inside the proton, which reflects the 
nonperturbative contributions. They produce the spin-flip 
effects in the pomeron coupling which do not vanish at 
high-energies. A similar form of the proton-pomeron coupling has 
been found in \cite{mod}. 

When the absolute values of initial and final proton momenta do 
not coincide, the functions $A$ and $B$ in (\ref{ver}) should 
depend on the fraction of the initial proton momentum $x_p$ 
carried off by the pomeron. It can be seen that this structure is 
connected with the non-forward gluon distributions. The spin-independent function $B$ might be related to $G(r,x_p)$  and spin-dependent part $A$ should be expressed in terms of $\Delta G(r,x_p)$ (see e.g. \cite{mank}). For zero $x_p$ they 
represent some proton form factors.

A convenient tool to study the spin-dependent pomeron structure 
might be the polarized diffractive leptoproduction reactions. We 
shall consider here the $A_{ll}$ asymmetry of $J/\Psi$ and $Q 
\bar Q$ production in these processes. It has been shown in 
\cite{gola_ll} that the $A_{ll}$ asymmetry in the diffractive 
processes is proportional to the fraction of the initial proton 
momentum $x_p$ carried off by the pomeron. The mass of the 
produced hadron system is determined by $M^2_h \sim s y x_p$. For  
diffractive  $J/\Psi$ production $M_h$ is a vector meson mass and 
we find that $x_p \sim (m_J^2+Q^2+|t|)/(s y)$. As a result, the 
relevant $A_{ll}$ asymmetry should decrease with growing energy. 
For the pomeron coupling  (\ref{ver}) spin asymmetry looks as 
follows 
\begin{equation} 
A_{ll} \sim \frac{|t|}{s}\frac{(2- 
y)(1+2 m \alpha_{flip})} {(2-2 y+ y^2)((1+2 m 
\alpha_{flip})^2+\alpha_{flip}^2 |t|)}, 
\end{equation} 
where 
$\alpha_{flip}=A(t)/B(t)$. Thus, the form of $A_{ll}$ asymmetry depends 
on the ratio of spin-flip to non-flip  parts of the pomeron 
coupling 
which have been found in \cite{gol_kr} to be about 0.1. The 
predicted asymmetry at HERMES energies is shown in Fig.1. At 
HERA energy, the asymmetry will be negligible. The $A_{ll}$ 
asymmetry might be connected with the spin-dependent gluon 
distribution $\Delta G$ only for $|t|=0$. We have found  that 
this asymmetry is equal to zero in the forward direction and 
$\Delta G$ can not be extracted from $A_{ll}$ in agreement with 
results of  \cite{mank}. However, this asymmetry might be 
expressed in terms of the non-forward gluon distributions inside 
the proton.

In the case of $Q \bar Q$ diffractive leptoproduction, the 
produced hadron mass is not fixed and $x_p$ is arbitrary,
typically, of about $.05-.1$. The $A_{ll}$ asymmetry in this case 
is proportional to $x_p$ as previously and it should have a weak 
energy dependence. This was confirmed by specific 
calculations. The cross section integrated over 
the pomeron momentum transfer was calculated because 
the recoil proton is usually not detected in diffractive 
experiments. Strong dependence of the polarized
cross sections on the mass of produced quarks and $\beta 
\simeq Q^2/(Q^2+M_x^2)$ has been observed. Some results for 
the spin-dependent vertex and forms of the cross 
sections can be found in \cite{gola_ll}. The estimated 
$A_{ll}$ asymmetry of the diffractive open charm ($c \bar c$) 
production  is shown in Fig.2.  

The standard pomeron coupling with the pomeron looks as $\bar 
u(p_2) \gamma_{\mu} u(p_1)$, which is completely equivalent to the 
form of lepton-photon interaction. This leads to the same 
spin-dependent parts in the lepton and hadron blocks and a non small amount of $A_{ll}$  asymmetry for $\alpha_{flip}=0$ 
(see Figs. 1 and 2). This asymmetry might be an important 
background in the lepton-proton experiments where the final 
proton is not detected. The predicted non small value of the
asymmetry in diffractive $Q \bar Q$ production and its weak 
energy dependence permit one to study, in these reactions, the 
spin-dependent gluon distributions inside the proton at $\sqrt{s} 
\leq 20 \mbox{GeV}^2$ as well as at HERA energies. \\
\begin{minipage}{6.cm}
\epsfxsize=6.cm
\epsfbox{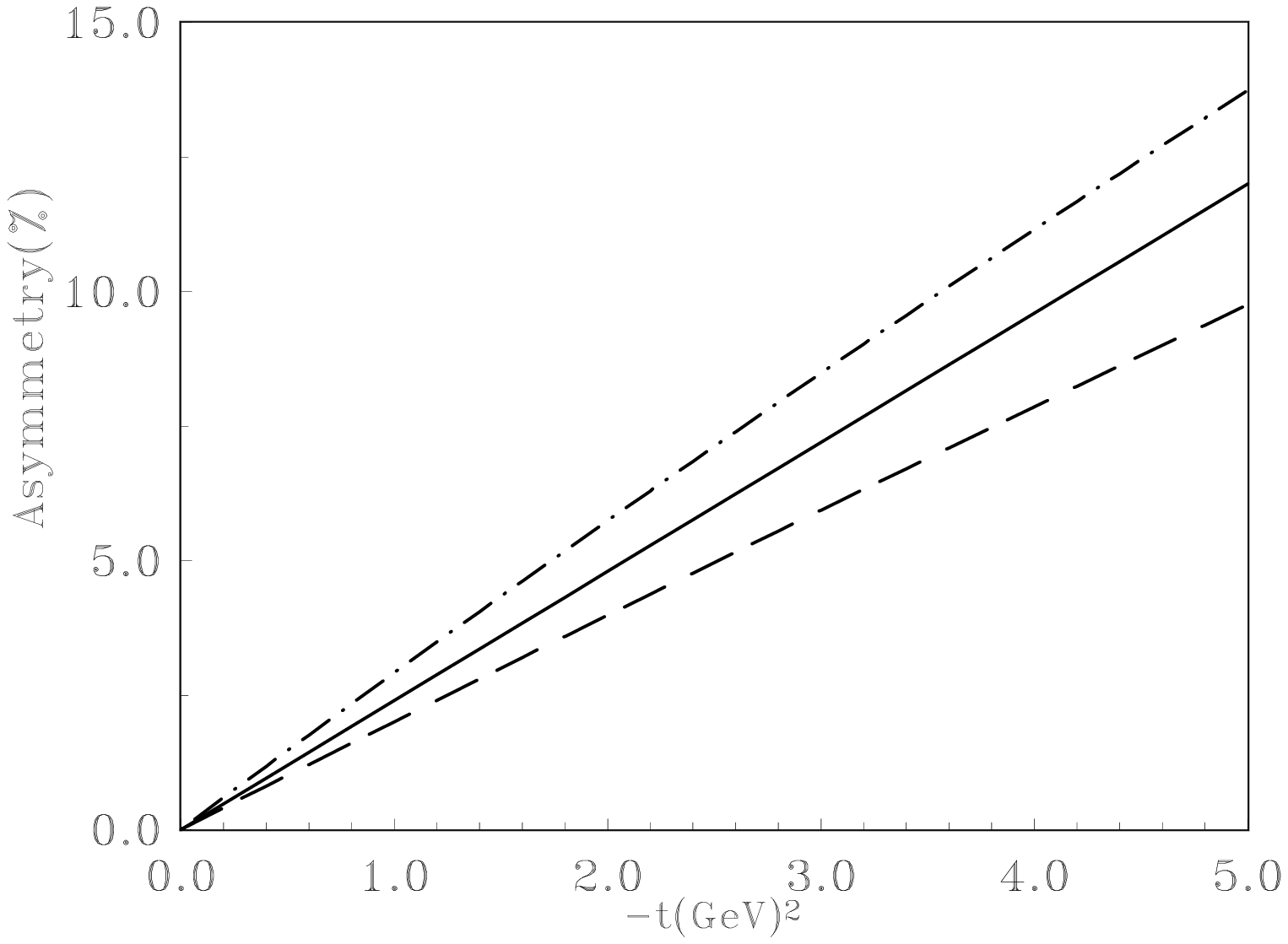}
\end{minipage}
\begin{minipage}{0.cm}
\end{minipage}
\begin{minipage}{6.cm}
\epsfxsize=6.cm
\epsfbox{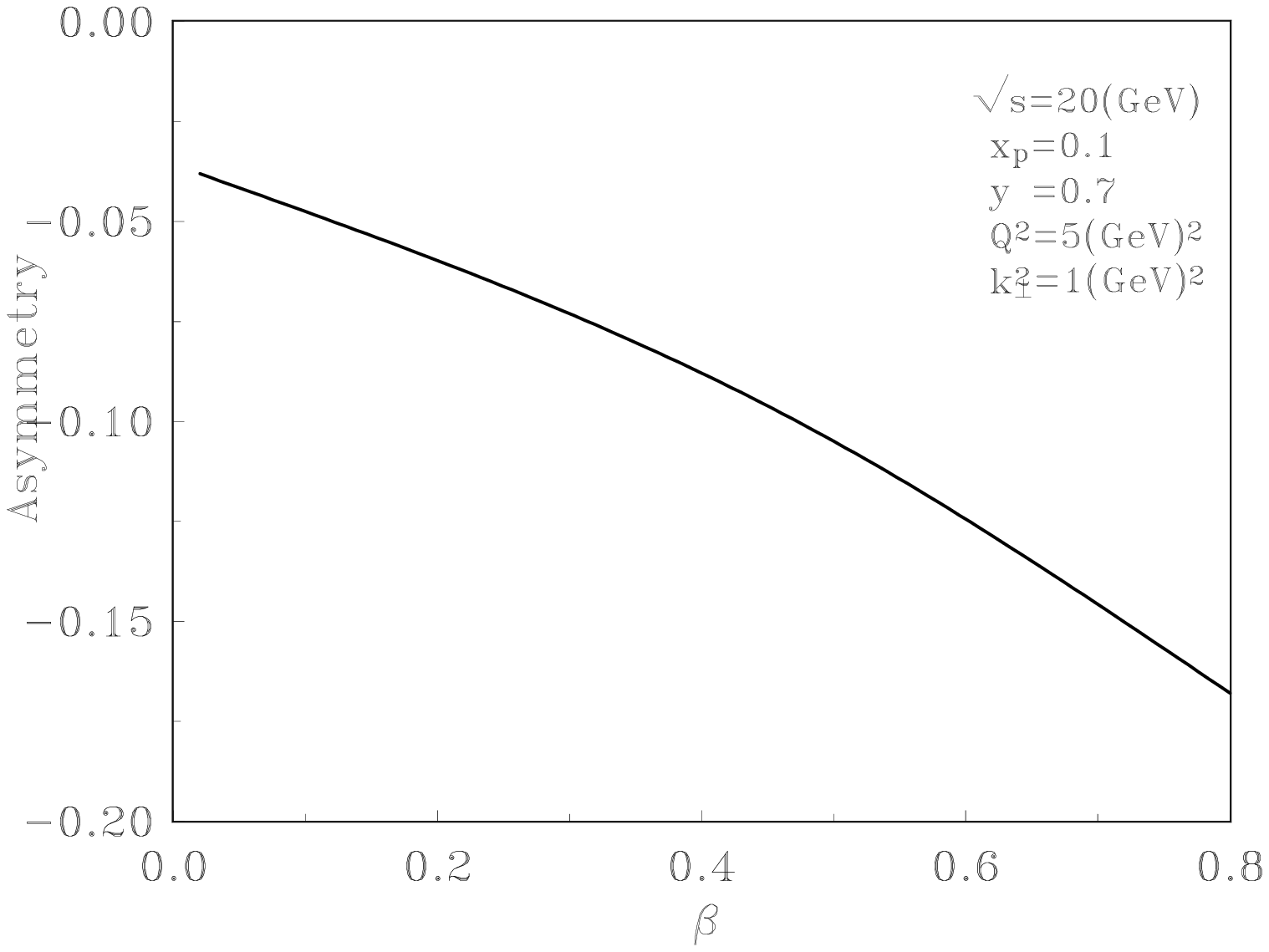}
\end{minipage}
\begin{minipage}{5.7cm}
Fig.1~ $A_{ll}$ asymmetry of $J/\Psi$ production  at HERMES:
solid line -for $\alpha_{flip}=0$; dot-dashed line -for 
$\alpha_{flip}=-0.1$;
dashed line -for $\alpha_{flip}=0.1$.
\end{minipage}
\begin{minipage}{.3cm}
\phantom{aaa}
\end{minipage}
\begin{minipage}{5.5cm}
Fig.2~$\beta$-- dependence of $A_{ll}$ asymmetry of diffractive
open charm production for the standard  pomeron coupling.
\end{minipage}

\bigskip

We have found that the spin structure of the pomeron coupling 
should modify the spin--dependent cross section in diffractive 
processes. Not small values of the $A_{ll}$ asymmetry in the 
diffractive $Q \bar Q$ production have been predicted. The 
asymmetry is free from the normalization factors and is sensitive 
to the dynamics of pomeron interaction. Thus, the $A_{ll}$ 
asymmetry in diffractive $J/\Psi$ and $ Q \bar Q$ leptoproduction 
is convenient to test the pomeron coupling structure and the 
non-forward spin-dependent gluon distributions. The polarized 
diffractive $\phi$ --meson leptoproduction can be used for this purpose too.

\end{document}